\documentclass[aps,prb,twocolumn,showpacs,superscriptaddress]{revtex4-1}
\usepackage{graphicx}
\usepackage{amsmath}
\usepackage{amssymb}
\usepackage{color}
\usepackage[usenames,dvipsnames]{xcolor}
\usepackage{comment}
\usepackage[colorlinks=true,linkcolor=Maroon,citecolor=OliveGreen,urlcolor=Blue,linktoc=page]{hyperref}

\newcommand{\ocite}[1]{[\onlinecite{#1}]}
\newcommand{\sgn}{\operatorname{sgn}}
\renewcommand{\Re}{\mbox{Re} }
\renewcommand{\Im}{\mbox{Im} }
\newcommand{\Em}{\mathcal{E}_{m}}

\def\XXint#1#2#3{{\setbox0=\hbox{$#1{#2#3}{\int}$ }
\vcenter{\hbox{$#2#3$ }}\kern-.5\wd0}}

\newcommand*{\pvint}
{
\mathop{\,\,\vphantom{\intop}\!\!\!
\mathpalette\pvintop\relax}\nolimits}
\newcommand*{\pvintop}
[2]{
\ooalign{$#1\intop$\cr\hidewidth$#1-$\hidewidth}}
\newcommand*{\PVI}
{\pvint}

\begin{document}

\title{Effect of Fluctuations on the NMR Relaxation Beyond the Abrikosov Vortex State}
\author{A. Glatz}
\affiliation{Materials Science Division, Argonne National Laboratory,
	9700 S. Cass Avenue, Argonne, Illinois 60639, USA} 
\affiliation{Department
	of Physics, Northern Illinois University, DeKalb, Illinois 60115, USA}
\author{A. Galda}
\affiliation{Materials Science Division, Argonne National Laboratory,
	9700 S. Cass Avenue, Argonne, Illinois 60639, USA}
\author{A.A. Varlamov}
\affiliation{CNR-SPIN, Viale del Politecnico 1, I-00133, Rome, Italy} 
\affiliation{Materials Science Division, Argonne National Laboratory,
	9700 S. Cass Avenue, Argonne, Illinois 60639, USA}

\date{\today}

\begin{abstract}
The effect of fluctuations on the nuclear magnetic resonance (NMR)
relaxation rate $W=T_{1}^{-1}$ is studied in a complete phase diagram of a
2D superconductor above the upper critical field line $H_{\mathrm{c2}}(T)$ . In the
region of relatively high temperatures and low magnetic fields, the
relaxation rate $W$ is determined by two competing effects. The first one is
its decrease in result of suppression of quasi-particle density of states
(DOS) due to formation of fluctuation Cooper pairs (FCP). The second one is
a specific, purely quantum, relaxation process of the Maki-Thompson (MT)
type, which for low field leads to an increase of the relaxation rate. The
latter describes particular fluctuation processes involving self-pairing of
a single electron on self-intersecting trajectories of a size up to
phase-breaking length $\ell _{\phi }$ which becomes possible due to an
electron spin-flip scattering event at a nucleus. As a result, different
scenarios with either growth or decrease of the NMR relaxation rate are
possible upon approaching the normal metal -- type-II superconductor
transition. The character of fluctuations changes along the line $H_{\mathrm{c2}}(T)$
from the thermal long-wavelength type in weak magnetic fields to the
clusters of rotating FCP in fields comparable to $H_{\mathrm{c2}}(0)$. We find that
below the well-defined temperature $T^{\ast }_0\approx 0.6T_{\mathrm{c0}}$, the MT
process becomes ineffective even in absence of intrinsic pair-breaking.  The
small scale of FCP rotations ($\xi _{\rm{xy}}$)  in so high fields impedes
formation of long ($\lesssim \ell _{\phi }$) self-intersecting trajectories,
causing the corresponding relaxation mechanism to lose its efficiency. This
reduces the effect of superconducting fluctuations in the domain of high
fields and low temperatures to just the suppression of quasi-particle DOS,
analogously to the Abrikosov vortex phase below the $H_{\mathrm{c2}}(T)$ line. 
\end{abstract}

\maketitle


\section{Introduction}

Nuclear magnetic resonant (NMR) spin-lattice relaxation is a result of
nuclei-interactions with low frequency excitations available in the
investigated system\cite{S90}. This fact makes NMR a powerful tool for
studying low-energy excitation dynamics in novel materials\cite{RBK98}.

In the Abrikosov phase of the type-II superconductors, for magnetic fields
well above the critical field $H_{\mathrm{c1}}$ but still below $H_{\mathrm{%
c2}}$, magnetic flux lines are separated by superconducting regions at
distances of the order of the coherence length $\xi _{\rm{xy}}$. The low-energy
excitations driving spin-lattice relaxation are the weighted average of the
intra-vortex excitations and of the contribution from the inter-vortex
regions, possibly connected by a spin diffusion process\cite{S90}. In the
vortex liquid phase, flux line diffusion provides an additional possible
relaxation mechanism (see Ref.~[\onlinecite{C97}] and references therein).

In a recent work\cite{GVV11a} the authors pointed out that a dynamic state
with clusters of coherently rotating FCP is formed above the $H_{\mathrm{c2}%
}(T)$ line at low temperatures. It is therefore of special interest to study
the effect of this fluctuation analogue of the vortex state on the magnetic
field dependence of the relaxation rate near $H_{\mathrm{c2}}\left( T\right) 
$. Some preliminary experimental studies were performed\cite{LR} by
measuring the $^{11}B$ NMR relaxation rates in a single crystal of
superconducting Y\!Ni$_{2}$\!B$_{2}$ (${T_{\mathrm{c0}}=15.3}$K in zero field).
The authors discussed an anomalous peak in the NMR relaxation rate magnetic
field dependence $W(H)$ at temperatures $2K$ and $4K$ in fields close to $H_{%
\mathrm{c2}}(T)$, which they tentatively attributed to quantum fluctuations
of magnetic flux lines.

\begin{figure}[th]
\includegraphics[width=\columnwidth]{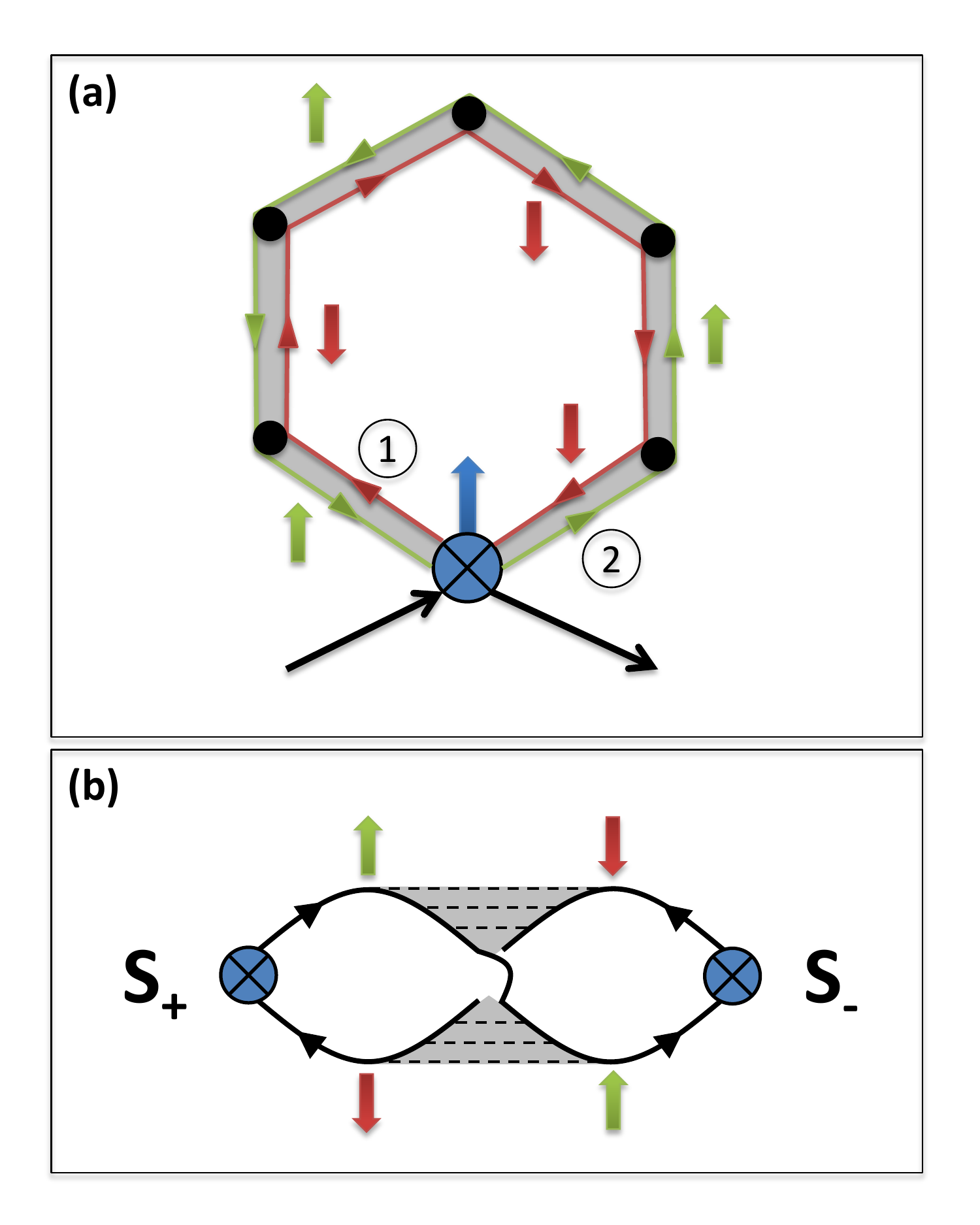}
\caption{(Color online) The relaxation process of Maki-Thompson (MT) type
related to the fluctuation pairing of electrons on self-intersecting
trajectories involving their spin-flip scatterings on the investigated
nuclei. \textbf{(a)} Initially electron moves along the trajectory 1 (clockwise, red),
due to several impurity scattering events it returns to the departure point.
In result of interaction with the nuclei, its spin and momentum flip and
electron returns along almost the same trajectory 2 (counter-clockwise, green). During this
motion it interacts with \textquotedblleft itself in past\textquotedblright\
(corresponding superconducting interactions are shaded). Such
process is possible due to \textquotedblleft fast\textquotedblright\ motion
of the electron along the trajectory and retarded character of
electron-phonon interaction. \textbf{(b)} Representation of the same process
as a Feynman diagram (compare with Fig.~\ref{fig.all_dia}(c).}
\label{fig.trajectory}
\end{figure}

Superconducting fluctuations affect the NMR spin-lattice relaxation rate of
superconductors in a wide range of magnetic fields and temperatures above
the upper critical field line $H_{\mathrm{c2}}(T)$~[%
\onlinecite{MA76, KF89, H92,
RV94, ERS99, MRV00,exp1,exp2,exp3,exp4,exp5}]. First of all, they suppress
the density of quasi-particle excitations\cite{ARW70, CCRV90}, which enters
quadratically into the NMR relaxation rate, and, as a consequence, they
reduce $W$. Nevertheless, this is not the only way for fluctuations to
influence nuclear relaxation. There exists another, purely quantum,
relaxation process of the Maki-Thompson (MT) type which consists of the
fluctuation self-pairing of a single electron on a self-intersecting
trajectory due to an electron spin-flip scattering event at a particular
nucleus\cite{MA76, KF89, LV09} (see Fig.~\ref{fig.trajectory}). The latter
process opens a new channel of NMR signal relaxation leading to the increase
of $W$.

Below, intending to investigate first of all the region of the Abrikosov
lattice formation from the side of normal metal, we study the effect of
superconducting fluctuations both of the thermal and the quantum nature on
the NMR relaxation mechanisms. We concentrate mainly on the most interesting
case of a two-dimensional $s-$wave superconductor restricting our
consideration by the representative dirty limit ${T_{\mathrm{c0}}\tau \ll 1}$,
where $\tau $ is the electron scattering time. We will derive the general
expression for the fluctuation contribution to the NMR\ relaxation rate
valid for the whole phase diagram above the line $H_{\mathrm{c2}}(T)$. Its
analysis for temperatures close to $T_{\mathrm{c0}}$ and fields much less
than $H_{\mathrm{c2}}(0)$ confirms the picture of the competition of two
contributions already studied in previous theoretical works\cite{MA76, KF89,
H92, RV94, ERS99, MRV00}.

The situation\ qualitatively changes when temperature decreases well below $%
T_{\mathrm{c0}}$. The nontrivial finding consists of the fact that below
some universal temperature ${T_{\rm{2D,0} }^{\ast}\simeq 0.6T_{\mathrm{c0}}}$ the superconducting fluctuations are no longer able  to
contribute positively to the NMR relaxation: the coherent MT scattering is
suppressed by strong fields ${H\gtrsim H_{\mathrm{c2}}(T_{ \rm{2D}
}^{\ast })}$, and the remaining effect of the quasi-particle DOS depletion
results in the opening of a fluctuation spin gap in the magnetic field
dependence of $W$. In order to compare the obtained results with the
available low-temperature experimental data\cite{LR,LRZ05}, in the last section we
extend our theory to the case of quasi-two-dimensional spectrum and study
the evolution of the crossover temperature $T^{\ast }_0$ (in general, $T^{\ast}$ is a function of the pair-breaking parameter and we denote the lowest value of $T^{\ast}$ for vanishing pair-breaking as $T^{\ast}_0$) versus the anisotropy
parameter $r=4\xi _{z}^{2}/\xi _{\rm{xy}}^{2}$ of a layered superconductor. It
turns out that three-dimensialization of the spectrum increases the value of 
$T^{\ast }$ with respect to $T_{\rm{2D} }^{\ast },$ \ which
completely excludes the superconducting quantum fluctuations as the reason
of the peak in NMR relaxation rate observed in Ref.~[\onlinecite{LR}] for ${T<0.25T_{\mathrm{c0}}}$.

The paper is organized as follows. In Sec.~II we introduce the method of calculating superconducting fluctuation corrections to the NMR relaxation rate. The only relevant contributions to the first order in perturbation theory, the DOS and MT processes, are calculated separately in Sections~III and~IV, respectively. In Section~V we present the total correction to the normal metal Korringa law, which is the main result of this paper. We derive asymptotic expressions for the total NMR correction in the regimes of quantum and thermal fluctuations and provide a rigorous numerical analysis of the results. In Section~VI we present the generalization of the approach to quasi-2D and 3D superconducting materials and outline the main consequences of the above generalization. Finally, Section VII summarizes the main results of the paper and explains the physical picture behind the competition of the DOS and MT relaxation processes at different temperatures and magnetic fields in the fluctuation regime. The crossover temperature between the two regimes is obtained from qualitative considerations.

\section{Model}

We begin with the dynamic spin susceptibility ${\chi _{\pm }^{R}(\mathbf{k},\omega )=\chi _{\pm }(\mathbf{k},\omega _{\nu }\rightarrow -i\omega +0^{+})}$, where 
\begin{equation}
\chi _{\pm }(\mathbf{k},\omega _{\nu })\!=\!\int_{0}^{1/T}\!\!\!\!d\varsigma
e^{i\omega _{\nu }\varsigma }\!\left\langle \hat{T}_{\varsigma }\!\left( 
\hat{S}_{+}(\mathbf{k},\varsigma )\hat{S}_{-}(-\mathbf{k},0)\right)
\!\right\rangle .
\end{equation}%
Here $\hat{S}_{\pm }$ are the spin raising and lowering operators, $%
\varsigma $ is the imaginary time, $\hat{T}_{\varsigma }$ is the time
ordering operator, $\mathbf{k}$\ is momentum, $\omega _{\nu }=2\pi T\nu $ ($%
\nu =0,1,2...$) is a bosonic Matsubara frequency corresponding to the
external field, and the angle brackets denote thermal and impurity averaging
in the usual way. In what follows we use the system of units where $\hslash
=k_{B}=c=1.$ The NMR relaxation rate $W$ is determined by the imaginary part
of the static limit of the dynamic spin susceptibility integrated over all
momenta: 
\begin{equation}
W=T\lim_{\omega \rightarrow 0}\frac{A}{\omega }\,\Im \!\int \left( d\mathbf{k%
}\right) \chi _{+-}^{R}(\mathbf{k},-i\omega )\,,  \label{korringa}
\end{equation}%
where $\left( d\mathbf{k}\right) \equiv d^{D}k/(2\pi )^{D},D$ is the
spectrum dimensionality, $A$ is a positive constant involving the
gyromagnetic ratio.

For noninteracting electrons $\chi _{\pm }^{\left( 0\right) }(\mathbf{k}%
,\omega _{\nu })$ is determined by the correlator of two single-electron
Green's functions ${G\left( \mathbf{k},\varepsilon _{n}\right) =\left( i%
\widetilde{\varepsilon }_{n}-\xi \left( \mathbf{k}\right) \right) ^{-1}}$, [${\widetilde{\varepsilon }_{n}=\varepsilon _{n}+\left( 2\tau \right) ^{-1}%
\mathrm{sign}\!\left( \varepsilon _{n}\right)}$, ${\varepsilon _{n}=2\pi T(
n+1/2) }$ is the fermionic Matsubara frequency$,\xi \left( \mathbf{k}%
\right) $ is the quasiparticle energy measured from the Fermi level], i.e.
by the usual loop diagram with the $\hat{S}_{\pm }(\mathbf{k},\tau )$
operators playing the role of external vertices (electron interaction with
the external field). Its trivial calculation leads to the well-known
Korringa law: ${W_{0}=4\pi ATN^{2}(0)}$, with $N(0)$\ as the one-electron
density of states. Below we will present the fluctuation contribution to $W$
in the dimensionless form by normalizing it to the latter result.

\begin{figure}[tbp]
\includegraphics[width=1.0\columnwidth]{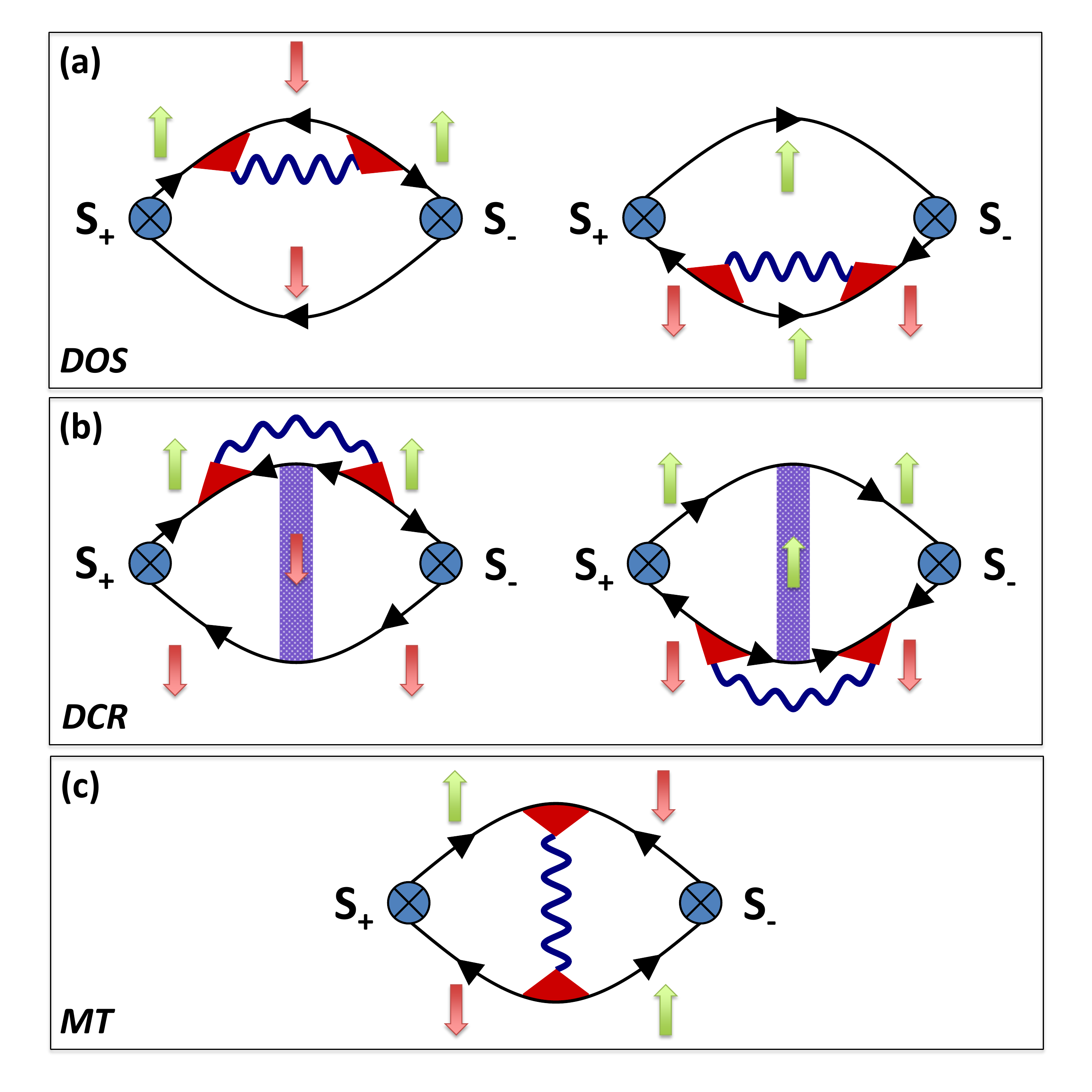}
\caption{(Color online) The diagrams for spin susceptibility. The solid
lines correspond to free electron Green's functions, bold wavy lines to the
fluctuation propagator, dashed triangles and rectangles account for
electrons scatterings on impurities. The two diagrams \textbf{(a)} represent
the density of states (DOS) correction, the diagrams \textbf{(b)} represent
the renormalization of the diffusion coefficient (DCR), while the diagram 
\textbf{(c)} corresponds to the Maki-Thompson (MT) process.}
\label{fig.all_dia}
\end{figure}

The first-order fluctuation contributions to $\chi _{\pm }$ in a dirty
superconductor above the line $H_{\mathrm{c2}}(T)$ can be expressed by means
of the standard fluctuation \textquotedblleft dressing\textquotedblright\ {%
(see Fig.~\ref{fig.all_dia})} of the loop of two Green's functions by the
fluctuation propagator $L_{m}(\Omega _{k})$ (wavy lines in the diagrams) and
impurity vertices $\lambda _{m}$ and $C_{m}$ (shaded three- and four-leg
blocks representing the result of averaging of the two Green's functions
products over elastic impurity scatterings{\ in the ladder approximation}).
Their explicit expressions in the representation of Landau levels and
Matsubara frequencies read as 
\begin{align}
& L_{m}^{-1}(\Omega _{k})=  \label{propagator} \\
& -N(0)\left\{ \ln \frac{T}{T_{\mathrm{c0}}}+\psi \left[ \frac{1}{2}+\frac{%
|\Omega _{k}|+\omega _{\mathrm{c}}\left(m+\frac{1}{2}\right)}{4\pi T}\right] -\psi
\left( \frac{1}{2}\right) \right\}   \notag
\end{align}%
\begin{equation}
\lambda _{m}(\varepsilon _{1},\varepsilon _{2})=\frac{\tau ^{-1}\Theta
(-\varepsilon _{1}\varepsilon _{2})}{|\varepsilon _{1}-\varepsilon
_{2}|+\omega _{\mathrm{c}}(m+1/2)+\tau _{\phi }^{-1}},  \label{lambda}
\end{equation}%
and the four-leg Cooperon is ${C_{m}=\left[ 2\pi N(0)\tau \right]
^{-1}\lambda _{m}}$. Here $m$ is the quantum number of the FCP Landau state, ${\Omega _{k}=2\pi Tk}$ (${k=0,\pm 1,\pm 2...)}$ is the bosonic Matsubara
frequency corresponding to the FCP, $\Theta (x)$ is the Heaviside theta
function.  An important characteristic of these expressions is that they are
valid even far from the critical temperature [for temperatures ${T\ll \min
\{\tau ^{-1},\omega _{D}\}}$] and for magnetic fields as strong as ${H\ll H_{\mathrm{c2}}(0) /( T_{\mathrm{c0}}\tau)}$. For the
sake of convenience, we introduce the reduced temperature $t=T/T_{\mathrm{c0}%
}$ and reduced magnetic field 
\begin{equation*}
h=\frac{\pi ^{2}}{8\gamma _{E}}\frac{H}{H_{\mathrm{c2}}(0)}=0.69%
\frac{H}{H_{\mathrm{c2}}(0)},
\end{equation*}%
where ${\gamma _{E}\simeq 1.78}$ is the Euler constant. The propagator~(\ref%
{propagator}) in these variables takes the form ${L_{m}^{-1}(T,H,\Omega
_{k})}={-N(0)\mathcal{E}_{m}( t,h,|k|) }$, with 
\begin{equation}
\mathcal{E}_{m}\left( t,h,x\right) =\ln t+\psi \left[ \frac{1+x}{2}+\frac{2h%
}{t}\frac{\left( 2m+1\right) }{\pi ^{2}}\right] -\psi \left( \frac{1}{2}%
\right) .  \label{Em}
\end{equation}%
We will also use its derivatives ${\mathcal{E}_{m}^{(p)}\left( t,h,x\right)}
\equiv {\partial _{x}^{p}\mathcal{E}_{m}\left( t,h,x\right) }$, which can be
expressed through polygamma functions: 
\begin{equation}
2^{p}\mathcal{E}_{m}^{(p)}\left( t,h,x\right) =\psi ^{\left( p\right) }\left[
\frac{1+x}{2}+\frac{2h}{t}\frac{\left( 2m+1\right) }{\pi ^{2}}\right] \,.
\label{poly}
\end{equation}%

Let us return to Fig.~\ref{fig.all_dia}. The two diagrams (a) represent the
effect of fluctuations on the single-particle self-energy, leading to a
decrease in corresponding DOS at the Fermi level. Consequently, in
accordance with the Korringa law, one can expect them to reduce the
relaxation rate $W$ with respect to its normal value, opening some kind of
fluctuation spin-gap upon approach of the transition line $H_{\mathrm{c2}%
}(T) $ from the normal phase.

Diagrams (b) with the four-leg Cooperon impurity blocks account for the
corrections to the NMR relaxation rate due to the electron diffusion
coefficient renormalization (DCR) by superconducting fluctuations. The
analogous contribution turns out to be the dominant one in the region of
quantum fluctuations in the case of fluctuation conductivity\cite{GVV11b}.
However, in the case under consideration, the additional integration over
the external momentum with respect to the case of conductivity makes their
contribution proportional to the square of the small Ginzburg-Levanyuk number\cite{LV09} 
\begin{equation*}
\rm{Gi}_{\rm{2D} }=\frac{7\zeta(3) }{32\pi ^{3}N(0)T_{\mathrm{c0}}\xi
^{2}}\,,
\end{equation*}%
which strongly suppresses the entire DCR contribution\cite{RV94}.

Finally, the diagram in Fig.~\ref{fig.all_dia}(c) is nothing else but the
diagrammatic representation of the MT process shown in Fig.~\ref%
{fig.trajectory}, where the region of attractive interaction (in grey)
interrupted periodically by impurity scattering events (circles) is replaced
by the fluctuation propagator (wavy line). This MT type diagram for $\chi
_{\pm }$ in this graphic form appears to be identical to the one for
conductivity. Nevertheless, the process shown in Fig.~\ref{fig.trajectory}%
(c) shows us the important difference in the topology of the former and
latter that arises from the spin structure. The MT diagram in Fig.~\ref%
{fig.all_dia}(c) is a non-planar graph with a single fermion loop. In
contrast, the MT graph for conductivity is planar and has two fermion loops.
The number of loops, according to the rules of the diagrammatic technique%
\cite{AGD}, determines the sign of the contribution. In the case of spin
susceptibility, which is under consideration, the topological sign of the MT
diagram turns out to be opposite to that one for conductivity.

The presence of the $\hat{S}_{\pm }(\mathbf{k},\tau )$ operators, taking
over the role of external vertices, changes not only the formal sign of the
MT diagram. The fact that two fermion lines attached to such vertex must
have the opposite spin labels (up and down) eliminates the Aslamazov-Larkin
diagram from our present consideration: one simply cannot consistently
assign a spin label to its central fermion lines for spin--singlet pairing%
\cite{MA76}.

\section{DOS contribution}

Let us start with the calculation of the DOS contribution determined by the
two diagrams in Fig.~\ref{fig.all_dia}(a). The corresponding expression for
the dynamic spin susceptibility integrated over all momenta is 
\begin{eqnarray}
&&\int \left( d\mathbf{k}\right) \chi _{+-}^{\rm{DOS}}(\mathbf{k},\omega _{\nu })=%
\frac{h}{\pi \xi ^{2}}\sum_{m}T\sum_{\Omega _{k}}L_{m}\left( \Omega
_{k}\right)   \notag \\
&&\cdot T\sum_{\varepsilon _{n}}\lambda _{m}^{2}\left( \varepsilon
_{n},\!\Omega _{k-n}\!\right) g\left( \varepsilon _{n+\nu }\!\right)
J^{\rm{DOS}}\!\left( \varepsilon _{n},\Omega _{k-n}\right) \,,  \label{DOS}
\end{eqnarray}%
where the first summation is performed over Landau levels and 
\begin{equation}
g\left( \varepsilon _{n+\nu }\right) =\int \left( d\mathbf{k}\right) G\left( 
\mathbf{k},\varepsilon _{n+\nu }\right) =-i\pi N(0)\sgn\left(
\varepsilon _{n+\nu }\right) \,,  \label{Gint}
\end{equation}%
\begin{align}
J^{\rm{DOS}}\left( \varepsilon _{n},\Omega _{k-n}\right) & =\int \left( d\mathbf{p%
}\right) ~G^{2}\left( \mathbf{p},\varepsilon _{n}\right) G\left( \mathbf{p}%
,\Omega _{k-n}\right)   \notag \\
& =2\pi iN(0)\frac{\Theta \left( -\varepsilon _{n}\Omega _{k-n}\right) 
\sgn\left( \varepsilon _{n}\right) }{\left( i\widetilde{\varepsilon }%
_{n}-i\widetilde{\Omega }_{k-n}\right) ^{2}}\,.  \label{Jint}
\end{align}%
In the approximation of a dirty metal (${T\tau \ll 1})$, 
\begin{eqnarray}
\int \left( d\mathbf{k}\right)\,  &&\chi _{+-}^{\rm{DOS}}(\mathbf{k},\omega _{\nu
})=\frac{2\pi N^{2}(0)T^{2}h}{\xi ^{2}}  \label{insuc} \\
&&\cdot \sum_{m}\sum_{\Omega _{k}}L_{m}\left( \Omega _{k}\right) \Xi
_{m}\left( \Omega _{k},\omega _{\nu }\right) \,  \notag
\end{eqnarray}%
with 
\begin{equation}
\Xi _{m}\left( \Omega _{k},\omega _{\nu }\right) =\sum_{\varepsilon _{n}}%
\frac{\Theta \left( -\varepsilon _{n}\Omega _{k-n}\right) \sgn\left(
\varepsilon _{n}\right) \sgn\left( \varepsilon _{n+\nu }\right) }{%
\left( \left\vert \varepsilon _{n}-\Omega _{k-n}\right\vert +\alpha _{m}%
\right) ^{2}}\,,  \label{sumge}
\end{equation}%
where we have defined ${\alpha _{m}\equiv \omega _{c}\left( m+1/2\right) }$.
For the Heaviside theta function we assume ${\Theta (0)=1}$. Now one can
perform the summation over fermionic frequencies by splitting its domain
into three intervals: ${(-\infty ,-\nu-1]},{[-\nu ,-1]},\text{ and }{[0,\infty )}$. The part of Eq.~(\ref{insuc}) depending on the external frequency $\omega
_{\nu },$ which determines the imaginary part of the susceptibility $\Im
\chi _{+-}^{R}(\mathbf{k},-i\omega )$ in Eq.~(\ref{korringa}), is 
\begin{equation}
\Xi _{m}^{\left( \omega \right) }\left( \Omega _{k},\omega _{\nu }\right)
=-2\sum_{n=0}^{\nu -1}\frac{\Theta \left( \varepsilon _{n}+\Omega
_{k}\right) }{\left( 2\varepsilon _{n}+\Omega _{k}+\alpha _{m}\right) ^{2}}%
\,.  \label{freqsum}
\end{equation}%
The summation over fermionic frequencies again can be performed by splitting
the domain of further summation over the bosonic frequencies into three: $%
k\in {\lbrack 0,\infty )},\,k\in {\lbrack -\nu ,-1]}$, and $k\in {(-\infty ,-\nu
-1]}$. Summation over the last interval results in zero and Eq.~(\ref{freqsum}%
) is presented as the sum of regular and anomalous parts: 
\begin{equation}
\Xi _{m}^{\left( \omega \right) }\left( \Omega _{k},\omega _{\nu }\right)
=\Xi _{m}^{\left( \mathrm{reg}\right) }\left( \Omega _{k},\omega _{\nu
}\right) +\Xi _{m}^{\left( \mathrm{an}\right) }\left( \Omega _{k},\omega
_{\nu }\right) \,,  \label{dossum}
\end{equation}%
where 
\begin{widetext}
\begin{equation}
\Xi _{m}^{\left( \rm{reg}\right) }\left( \Omega _{k},\omega _{\nu }\right) =\frac{%
\Theta \left( \Omega _{k}\right) }{8\pi ^{2}T^{2}}\left[ \psi ^{\prime
}\left( \frac{1}{2}+\frac{\omega _{\nu }}{2\pi T}+\frac{\Omega _{k}+\alpha
_{m}}{4\pi T}\right) -\psi ^{\prime }\left( \frac{1}{2}+\frac{\Omega
_{k}+\alpha _{m}}{4\pi T}\right) \right]  \label{regsum}
\end{equation}%
and 
\begin{equation}
\Xi _{m}^{\left( \rm{an}\right) }\left( \Omega _{k},\omega _{\nu }\right) =\frac{%
\Theta \left( -\Omega _{k}\right) \Theta \left( \Omega _{k}+\omega _{\nu
}\right) }{8\pi ^{2}T^{2}}\left[ \psi ^{\prime }\left( \frac{1}{2}+\frac{%
\omega _{\nu }}{2\pi T}+\frac{\Omega _{k}+\alpha _{m}}{4\pi T}\right) -\psi
^{\prime }\left( \frac{1}{2}+\frac{-\Omega _{k}+\alpha _{m}}{4\pi T}\right) %
\right] \,.  \label{ansum}
\end{equation}
\end{widetext}The regular part (\ref{regsum}) is an analytic function of the
external frequency $\omega _{\nu }$ and can be easily continued to the upper
half-plane of the complex frequencies by substitution ${\omega _{\nu
}\rightarrow -i\omega}$. As a result, by putting together Eqs.~(\ref%
{korringa}),(\ref{insuc}), (\ref{sumge}), and (\ref{regsum}), one finds [the
identity (\ref{poly}) was used to finalize $\Xi _{m}^{\left( \mathrm{reg}%
\right) }\left( \Omega _{k}\right) $]: 
\begin{equation}
W^{\rm{DOS}\left( \mathrm{reg}\right)}(t,h)=\frac{AN(0)}{8\pi ^{2}\xi ^{2}}%
h\sum_{m=0}^{M}\sum_{k=0}^{\infty }\frac{4\mathcal{E}_{m}^{\prime \prime
}\left( t,h,\left\vert k\right\vert \right) }{\mathcal{E}_{m}\left(
t,h,\left\vert k\right\vert \right) }\,,  \label{T_1reg}
\end{equation}%
with $M=\left( tT_{\mathrm{c0}}\tau \right) ^{-1}$ as the cut-off parameter.

\begin{figure}[th]
\begin{center}
\includegraphics[width=1.0\columnwidth ]{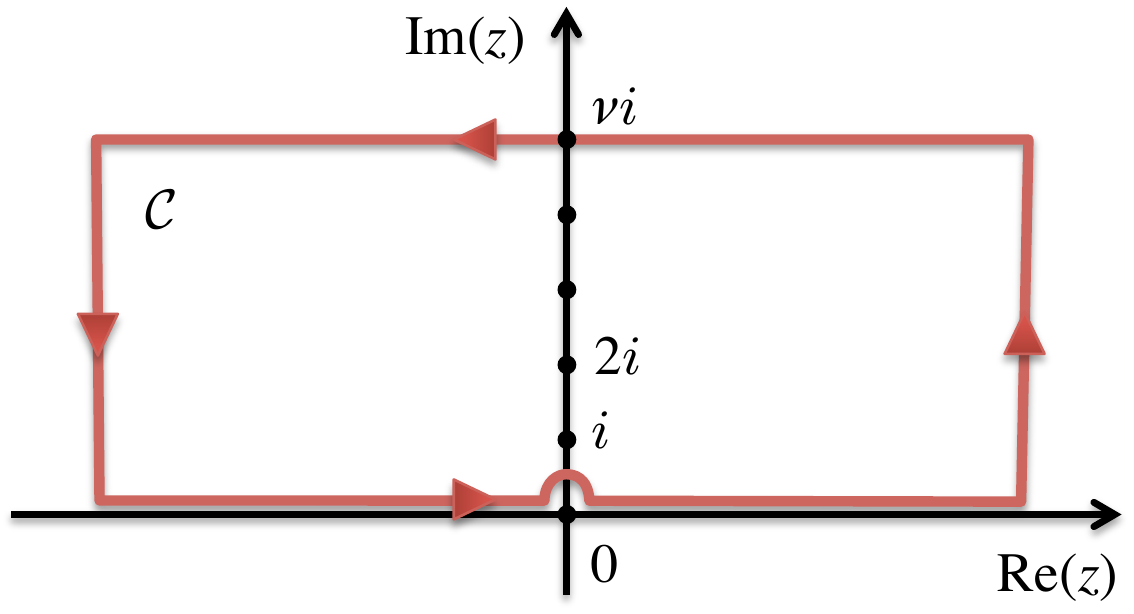}
\end{center}
\caption{(Color online) Closed integration contour $\mathcal{C}$ in the
plane of complex frequencies.}
\label{fig.mtcontour}
\end{figure}

Now, let us proceed to the analysis of the anomalous DOS contribution to the
relaxation rate determined by Eqs.~(\ref{korringa}), (\ref{insuc}), (\ref%
{sumge}), and (\ref{ansum}). Here, the upper limit of summation over bosonic
frequencies contains the external frequency and one should be cautious with
the analytic continuation to the upper half-plane of complex frequencies.
One can write 
\begin{equation}
\int \left( d\mathbf{k}\right) \chi _{+- }^{\rm{DOS}\left( \mathrm{an}\right)}(%
\mathbf{k},\omega _{\nu })=-\frac{N(0)h}{2\pi \xi ^{2}}\sum_{m}\theta
_{m}\left( \omega _{\nu }\right) \,,  \label{T_1an}
\end{equation}%
where 
\begin{equation}
\theta _{m}\left( \omega _{\nu }\right) =\sum_{l=1}^{\nu -1}f_{m}(l,\omega
_{\nu })  \label{theta}
\end{equation}%
and 
\begin{equation}
f_{m}(l,\omega _{\nu })=\frac{\left[ \mathcal{E}_{m}^{\prime }\left( 2\nu
-l\right) -\mathcal{E}_{m}^{\prime }\left( l\right) \right] }{\mathcal{E}%
_{m}\left( l\right) }\,.  \label{fm}
\end{equation}%
Note that the summation limit in Eq.~(\ref{theta}) can be extended from $%
{(\nu -1)}$ to $\nu $, since ${f_{m}(l=\nu ,\omega _{\nu })=0}$.

The analytic continuation of Eq.~(\ref{theta}) to the upper half-plane of
complex frequencies was performed in Ref.~[\onlinecite{AV80}] (see also
Eq.~(8.84) in Ref.~[\onlinecite{LV09}]). By means of the Eliashberg
transformation~\cite{E61}, the corresponding sum can be presented as an
integral over a counterclockwise closed contour $\mathcal{C}$ consisting of
two horizontal lines, two vertical lines, and a semicircle in the upper
complex plane around the pole ${z=0}$ (see Fig.~\ref{fig.mtcontour}): 
\begin{equation}
\theta _{m}\left( \omega _{\nu }\right) \!=\!\frac{1}{2i}\oint\limits_{%
\mathcal{C}}\coth \left( \pi z\right) f_{m}(-iz,\omega _{\nu })dz\,.
\label{eliash}
\end{equation}%
The integrals over the vertical line segments are zero and the integral over
the semi-circle reduces to minus half of the residue of the integrand at ${z=0}$. By inverting the direction of integration along the line ${\Im\, z=\nu }$
and shifting the integration variable as ${z-i\omega _{\nu }/2\pi
T\rightarrow z}$ in the corresponding integral, one finds: 
\begin{eqnarray}
&&\theta _{m}\left( \omega _{\nu }\right) =-\frac{f_{m}(0,\omega _{\nu })}{2}%
+\frac{1}{2i}\PVI_{-\infty }^{\infty }\coth \left( \pi z\right)  \notag \\
&&\times \left[ f_{m}(\!-iz,\omega _{\nu })\!-\!f_{m}(\!-iz\!+\!\omega _{\nu
}/2\pi T,\omega _{\nu })\right] dz\,.  \label{resi}
\end{eqnarray}%
Eq.~(\ref{resi}) is an analytic function of $\omega _{\nu }$ and one can
perform its continuation by the standard substitution ${\omega _{\nu
}\rightarrow -i\omega }$. By the change of variables ${z+\omega /2\pi
T\rightarrow z}$ in the second integral and with the help of the identity 
\begin{equation*}
\coth (a)-\coth (b)=-\frac{\sinh( a-b ) }{\sinh (a)\sinh (b)}\,,
\end{equation*}%
one finally finds 
\begin{align}
& \theta _{m}^{R}\left( -i\omega \right) =-\frac{f_{m}(0,-i\omega )}{2}
\label{thetar} \\
& +i\frac{\sinh \left( \omega /2T\right) }{2}\int_{-\infty }^{\infty }\frac{%
f_{m}(-iz,-i\omega )dz\!}{\sinh \left( \pi z\right) \sinh \left( \pi
z-\omega /2T\right) }\,.  \notag
\end{align}%
Substitution of the explicit expression for function ${f_{m}(-iz,-i\omega )}$
from Eq.~(\ref{fm}) into Eq.~(\ref{thetar}) results in 
\begin{widetext}
\begin{align}
\int \left( d\mathbf{k}\right) \chi _{+- }^{\rm{DOS}\left( \mathrm{an}\right)}(\mathbf{k},\omega ) =-\frac{N(0) h}{4\pi \xi ^{2}}\sum_{m=0}^{M}\left\{ -\frac{\mathcal{E}_{m}^{\prime }\left( -\frac{i\omega }{\pi T} \right) -\mathcal{E}_{m}^{\prime }( 0) }{\mathcal{E}_{m}( 0) }\right. \left. +i\sinh \left( \frac{\omega }{2T}\right) \PVI\limits_{-\infty}^{\infty }\frac{\left[ \mathcal{E}_{m}^{\prime }\left( -iz\!-\!\frac{i\omega}{\pi T}\right) \!-\!\mathcal{E}_{m}^{\prime }(-iz) \right]dz\!}{\sinh ( \pi z) \sinh \left( \pi z - \omega /2T\right) \mathcal{E}_{m}( -iz) }\right\}\,.  \label{KR2}
\end{align}
By sending the external frequency to zero, one finds the anomalous DOS contribution to the NMR relaxation rate:
\begin{eqnarray}
W^{\rm{DOS}(\rm{an})}(t,h) &=&\frac{A N(0)}{8\pi^{2}\xi ^{2}}h\sum_{m=0}^{M}\left\{ \frac{2\mathcal{E}_{m}^{\prime \prime }(0)}{\mathcal{E}_{m}( 0) } +2\pi \int_{-\infty}^{\infty }\frac{\Im\,\mathcal{E}_{m}^{\prime }( iz)\, \Im\,\mathcal{E}_{m}( iz)\, dz}{\sinh ^{2}( \pi z) \left[ \Re^{2}\mathcal{E}_{m}( iz) +\Im^{2}\mathcal{E}_{m}( iz)\right] } \right\}  \label{DOSanfin}
\end{eqnarray}
\end{widetext}Let us note that along the line $H_{\mathrm{c2}}(T),$ where the Eq.~(%
\ref{Em}) can be simplified as 
\begin{equation*}
\mathcal{E}_{m}\left( t,h,iz\right) =\frac{h-h_{\mathrm{c2}}( t) }{%
h_{\mathrm{c2}}( t) }+\frac{i\pi ^{2}zt}{4h_{\mathrm{c2}}( t) }\,,
\end{equation*}%
the second term in Eq.~(\ref{DOSanfin}) exactly cancels the first one and in
this region only the regular part of the DOS diagrams contributes to the NMR
relaxation rate. This fact justifies the approximation of static
fluctuations (account for the term with $\Omega _{k}=0$ only) assumed in the
previous works\cite{MA76, KF89, H92, RV94, ERS99, MRV00} in their
consideration performed close to $T_{\mathrm{c0}}.$

\section{Maki-Thompson contribution}

The mentioned above MT contribution to the process of NMR relaxation is
described by the diagram in Fig.~\ref{fig.all_dia}(c). The corresponding
expression for the dynamic spin susceptibility integrated over all momenta
is 
\begin{eqnarray}
&&\int \left( d\mathbf{k}\right) \chi _{+-}^{\rm{MT} }=\frac{hT^{2}%
}{2\pi \xi ^{2}}\sum_{m}\sum_{\Omega _{k}}L_{m}\left( \Omega _{k}\right)
\sum_{n}\lambda _{m}\left( \varepsilon _{n},\Omega _{k-n}\right)   \notag \\
&&\cdot \lambda _{m}\!\left( \varepsilon _{n+\nu },\Omega _{k-n-\nu }\right)
\!I_{2g}\!\left( \varepsilon _{n},\Omega _{k-n}\right) \!I_{2g}\!\left(
\varepsilon _{n+\nu },\Omega _{k-n-\nu }\right)  \label{MTint}
\end{eqnarray}%
Restricting ourselves by the assumed above case of the {dirty}
superconductor, one can write an explicit expression for the integral of the
product of two Green's functions: 
\begin{equation*}
I_{2g}\!=\!\!\int \!\!\left( d\mathbf{p}\right)\! G\!\left( \mathbf{p},\varepsilon
_{n}\right) \!G\!\left( \mathbf{-p},\Omega _{k-n}\right) \!=\!2\pi N(0)\tau
\Theta\!\left( -\varepsilon _{n}\Omega _{k-n}\right) 
\end{equation*}%
and express Eq.~(\ref{MTint}) in the standard form 
\begin{equation*}
\int \!\left( d\mathbf{k}\right) \chi _{+-}^{\rm{MT} }=\frac{2\pi
N^{2}(0)T^{2}h}{\xi ^{2}}\sum_{m}\sum_{\Omega _{k}}L_{m}\!\left( \Omega
_{k}\right) \Upsilon _{m}\!\left( \Omega _{k},\omega _{\nu }\right)\,.
\end{equation*}%
The summation over fermionic frequencies in the expression 
\begin{eqnarray*}
\Upsilon_{m} &=&\sum_{n}\frac{\Theta \left[ \varepsilon _{n}\left(
\varepsilon _{n}-\Omega _{k}\right) \right] \Theta \left[ \varepsilon
_{n+\nu }\left( \varepsilon _{n+\nu }-\Omega _{k}\right) \right] }{\left(
\left\vert 2\varepsilon _{n}-\Omega _{k}\right\vert +\alpha _{m}\right)
\left( \left\vert 2\varepsilon _{n+\nu }-\Omega _{k}\right\vert +\alpha
_{m}\right) } \\
&=&\Upsilon _{m}^{\left( \rm{reg1}\right) }\left( \Omega _{k},\omega _{\nu
}\right) +\Upsilon _{m}^{\left( \rm{reg2}+\rm{an}\right) }\left( \Omega _{k},\omega
_{\nu }\right) 
\end{eqnarray*}%
is performed in complete analogy with the previous section, and one finds as
a result: 
\begin{eqnarray*}
\Upsilon _{m}^{\left( \rm{reg1}\right) }\left( \Omega _{k},\omega _{\nu }\right) 
&=&\frac{1}{4\pi T}\frac{1}{\omega _{\nu }}\left[ \psi \left( \frac{1}{2}+%
\frac{2\omega _{\nu }+\left\vert \Omega _{k}\right\vert +\alpha _{m}}{4\pi T}%
\right) \right.  \\
&&\left. -\psi \left( \frac{1}{2}+\frac{\left\vert \Omega _{k}\right\vert
+\alpha _{m}}{4\pi T}\right) \right] \,
\end{eqnarray*}%
and%
\begin{eqnarray*}
&&\Upsilon _{m}^{\left( \rm{reg2}+\rm{an}\right) }\left( \Omega _{k},\omega _{\nu
}\right) =\frac{\Theta \left( \omega _{\nu -1}-|\Omega _{k}|\right) }{4\pi
T\left( \omega _{\nu }+\alpha _{m}\right) } \\
&&\cdot \left[ \psi \left( \frac{1}{2}+\frac{2\omega _{\nu }-|\Omega
_{k}|+\alpha _{m}}{4\pi T}\right) -\psi \left( \frac{1}{2}+\frac{|\Omega
_{k}|+\alpha _{m}}{4\pi T}\right) \right] \,.
\end{eqnarray*}%
Analitic continuation of $\Upsilon _{m}^{\left( \rm{reg1}\right) }\left( \Omega
_{k},\omega _{\nu }\right) $\ is trivial, while that one of $\Upsilon
_{m}^{\left( \rm{reg2}+\rm{an}\right) }$ is performed by means of the Eliashberg
transformation~(\ref{eliash}). Finally, one obtains 
\begin{widetext}
\begin{eqnarray}
W^{\rm{MT}}(t,h) &=&\frac{AN(0) }{16\pi ^{2}\xi ^{2}}\,h \sum_{m=0}^{M}%
\left[ \sum_{k=-\infty }^{\infty }\frac{4\mathcal{E}_{m}^{\prime \prime
}\left( t,h,\left\vert k\right\vert \right) }{\mathcal{E}_{m}\left(
t,h,\left\vert k\right\vert \right) }\right. \notag \\
&&\left. +\frac{\pi ^{3}}{%
\gamma _{\varphi }+\frac{2h}{t}( m+1/2) }\int\limits_{-\infty
}^{\infty }\frac{dz}{\sinh ^{2}( \pi z) } \frac{\Im^{2}\Em(
t,h,iz) }{\Re^{2}\Em( t, h, iz) + \Im^{2}\Em( t, h,
iz) }\right] \notag,
\end{eqnarray}
\end{widetext}with ${\gamma _{\phi }=\pi /( 8T_{\mathrm{c0}}\tau _{\phi
}) }$ and $\tau _{\phi }$ as the phase-breaking time.

\section{Main result}

Collecting DOS and MT contributions in one expression and normalizing it to
the normal metal Korringa relaxation rate, one can write the expression for $%
W^{\rm{fl}}$ valid in the whole phase diagram (with the restrictions discussed
above): 
\begin{widetext}
\begin{eqnarray}
\frac{W^{\rm{fl}}(t,h)}{W_0} &=&\frac{\rm{Gi}_{\rm{2D} }}{7\zeta( 3) }\left( \frac{h%
}{t}\right) \sum_{m=0}^{M}\left[ \sum_{k=-\infty }^{\infty }\frac{8\mathcal{E%
}_{m}^{\prime \prime }( t,h,\left\vert k\right\vert) }{\mathcal{E%
}_{m}( t,h,\left\vert k\right\vert) }\right. +4\pi \int_{-\infty
}^{\infty }\frac{dz}{\sinh ^{2}( \pi z) }\frac{\Im\, \mathcal{E}%
_{m}^{\prime }( t,h,iz)\, \Im\, \mathcal{E}_{m}( t,h,iz) }{%
\Re ^{2}\mathcal{E}_{m}( t,h,iz) +\Im ^{2}\mathcal{E}_{m}(t,h,iz) }  \notag \\
&&\left. +\frac{\pi ^{3}}{\gamma _{\phi }+\frac{2h}{t}( m+1/2) 
}\int\limits_{-\infty }^{\infty }\frac{dz}{\sinh ^{2}( \pi z) }%
\frac{\Im ^{2}\mathcal{E}_{m}( t,h,iz) }{\Re ^{2}\mathcal{E}%
_{m}( t,h,iz) +\Im ^{2}\mathcal{E}_{m}( t,h,iz) }%
\right] \,.  \label{TOT}
\end{eqnarray}%
\end{widetext}

One can analyze it in different limiting cases. \ Close to $T_{\mathrm{c0}}$
and for magnetic fields not too high (${h\ll 1}$) but arbitrary with respect
to reduced temperature ${\epsilon =\left( T-T_{\mathrm{c0}}\right) /T_{%
\mathrm{c0}}\ll 1}$ and phase-breaking rate ${\gamma _{\phi }\ll 1}$ one can
perform the integrations and summations in Eq.~(\ref{TOT}) and get 
\begin{widetext}
\begin{equation}
\frac{W^{\rm{fl}}( \epsilon ,h\ll 1)}{ W_0 } =-3\rm{Gi}_{\rm{2D} }\left\{ \left[
\ln \frac1h -\psi \left( \frac{\epsilon }{2h}+\frac{1}{2}\right) %
\right] -\frac{\pi ^{4}}{168\zeta( 3) }\frac{1}{\epsilon -\gamma
_{\phi }}\left[ \psi \left( \frac{\epsilon }{2h}+\frac{1}{2}\right) -\psi
\left( \frac{\gamma _{\phi }}{2h}+\frac{1}{2}\right) \right] \right\} . 
\label{general}
\end{equation}%
\end{widetext}In the limit of weak fields ${h\ll \min \{ \epsilon
,\gamma _{\phi }\} }$
\begin{eqnarray}
\frac{W^{\rm{fl}}}{W_{0}} &=&\,3\rm{Gi}_{\rm{2D} }\left[ \frac{\pi ^{4}}{%
168\zeta( 3) }\frac{1}{\epsilon -\gamma _{\phi }}\ln \frac{%
\epsilon }{\gamma _{\phi }}-\ln \frac{1}{\epsilon }\right]   \notag \\
&&-\rm{Gi}_{\rm{2D} }\,\frac{h^{2}}{2\epsilon ^{2}}\left[ \frac{\pi ^{4}}{%
168\zeta( 3) }\frac{\gamma _{\phi }+\epsilon }{%
\gamma _{\phi }^{2}}-1\right] .  \label{close}
\end{eqnarray}%
The first line of Eq.~(\ref{close}) reproduces the results of Refs.~\ocite%
{MA76,KF89,H92,RV94},\ while the magnetic field dependence of $W^{\rm{fl}}$ for
weak fields (second line of Eq.~(\ref{close})) was firstly analytically
found in Ref.~\ocite{MRV00}. One can see that the  MT contribution dominates
when the pair-breaking is weak. In this case superconducting fluctuations in
weak fields increase the NMR\ relaxation; increase of the field reduces the
latter. As the phase-breaking grows, the role of the first term in Eq.~(\ref%
{close}) weakens and the effect of fluctuations can change sign: the MT
trajectories shorten and the negative contribution of superconducting
fluctuations due to the suppression of the quasi-particle density of states
becomes the dominant. Since ${\gamma _{\phi }\lesssim 1}$ the effect of
magnetic field on $W^{\rm{fl}}$ is always negative.

In the opposite case ${1\gg h\gg \max \left\{ \epsilon ,\gamma _{\phi
}\right\} }$ the MT contribution dominates\cite{MRV00}: intrinsic pair-breaking
here is weak while the effect of magnetic field on the motion of Cooper
pairs is not yet strong enough:

\begin{equation}
\frac{W^{\rm{fl}}}{W_{0}}\approx \,3\rm{Gi}_{\rm{2D} }\left[ \frac{\pi ^{6}}{%
672\zeta( 3) }\frac{1}{h}-\ln \frac{1}{h}\right] .  \notag
\end{equation}%
Concluding discussion of the closeness of $T_{\mathrm{c0}}$, one can write
the explicit expression for $W^{\rm{fl}}$ along the line $H_{\mathrm{c2}}(T)$ in its
beginning, where ${\epsilon +h\rightarrow 0}$: 
\begin{equation}
\frac{W^{\rm{fl}}}{W_{0}}=\,3\rm{Gi}_{\rm{2D} }\left\{ \frac{\pi ^{4}}{%
168\zeta( 3) }\frac{2h}{( \epsilon +h) ( \gamma_{\phi }+h) }-\ln \frac{1}{h}\right\} .  \notag
\end{equation}

Now let us turn to the main subject of our study: the domain of the phase
diagram above the second critical field at relatively low temperatures. Our
general formula~(\ref{general}) allows to obtain the explicit analytical
expressions, for instance, along the line $H_{\mathrm{c2}}(T)$, where ${t\ll h_{\mathrm{c2}}(t)}$. Here the main contribution is due to the lowest Landau level of the FCP motion. Corresponding propagator~(\ref{propagator}) has the pole structure:%
\begin{equation*}
L_{0}^{R}( t,h,iz) =-\frac{1}{N(0)}\frac{1}{\widetilde{h}+\frac{%
i\pi ^{2}zt}{4h_{\mathrm{c2}}( t) }}.
\end{equation*}%
Performing summation over bosonic frequencies and integration in Eq.~(\ref{TOT}) one finds%
\begin{widetext}
	\begin{equation}
\frac{W^{\rm{fl}}\!\left(t\ll h_{\mathrm{c2}}(t)\right)}{ W_0 }=-\frac{4\pi ^{2}\rm{Gi}_{\rm{2D} }%
}{7\zeta( 3) }\left\{ \ln \frac{1}{\widetilde{h}}+\frac{2%
\widetilde{h}\gamma _{\phi }}{\pi ^{2}}\left[ \psi ^{\prime }\left( \frac{%
4h_{\mathrm{c2}}( t) \widetilde{h}}{\pi ^{2}t}\right) -\frac{\pi ^{2}t}{%
4h_{\mathrm{c2}}( t) \widetilde{h}}-\frac{1}{2}\left( \frac{\pi ^{2}t}{%
4h_{\mathrm{c2}}( t) \widetilde{h}}\right) ^{2}\right] \right\}  \notag
\end{equation}
\end{widetext}\ At very low temperatures ${t\ll \widetilde{h}}$, ${\widetilde{h}\equiv\left[ H-H_{\mathrm{c2}}(0)\right] /H_{\mathrm{c2}}(0)}$, and just above $H_{\mathrm{c2}}(0),$
the regime of quantum fluctuations is realized. They suppress the NMR\
relaxation due to decrease of the quasi-particle density of states. 
\begin{equation}
\frac{W^{\rm{fl}}\!\left( t\!\ll\! \widetilde{h}\right) }{W_{0}}=-\frac{4\pi
^{2}\rm{Gi}_{\rm{2D} }}{7\zeta( 3) }\!\left[ \ln \frac{1}{%
\widetilde{h}} \!+\! \frac{\pi ^{4}t^{3}\gamma _{\phi }}{192h_{\mathrm{c2}}^{3}(t) \widetilde{h}^{2}}\right]\,.\label{quant}
\end{equation}%
At higher temperatures, ${\widetilde{h}\ll t \ll h_{\mathrm{c2}}(t)}$, superconducting fluctuations become of thermal nature, while the DOS suppression of the NMR relaxation remains dominant: 
\begin{equation}
\frac{W^{\rm{fl}}\!\left( \widetilde{h} \!\ll\! t \!\ll\! h_{\mathrm{c2}}(t)\!\right) }{W_{0}} \!=\! -\frac{4\pi^{2}\rm{Gi}_{\rm{2D} }}{7\zeta( 3) }\!\!\left[ \ln \frac{1}{\widetilde{h}} \!+\! \frac{\pi ^{2}t^{2}\gamma _{\phi }}{16h_{\mathrm{c2}}^{2}(t) \widetilde{h}}\right]\,.\label{therm}
\end{equation}

\begin{figure}[tbp]
	\includegraphics[width=1.0\columnwidth]{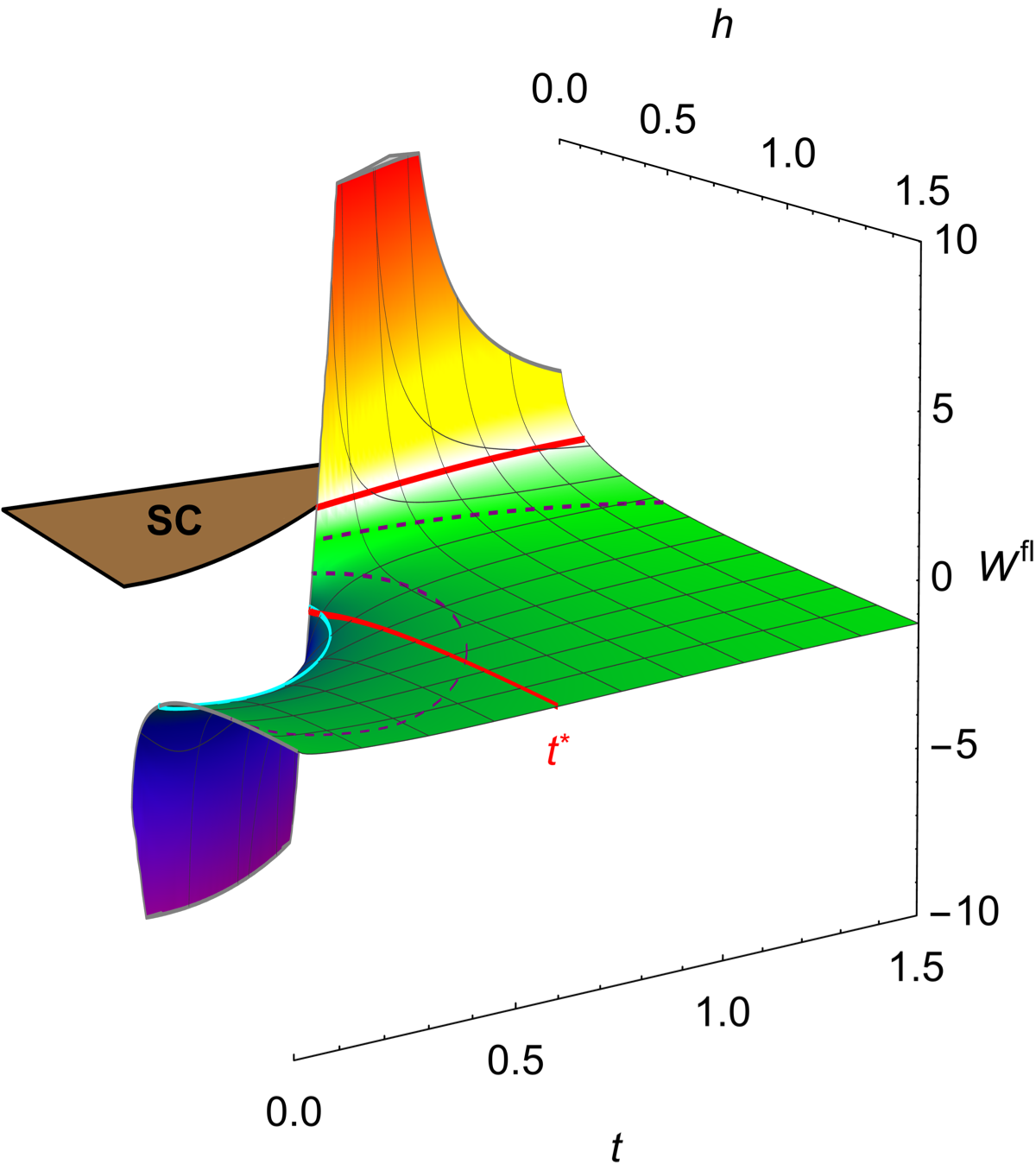}
	\includegraphics[width=0.75\columnwidth]{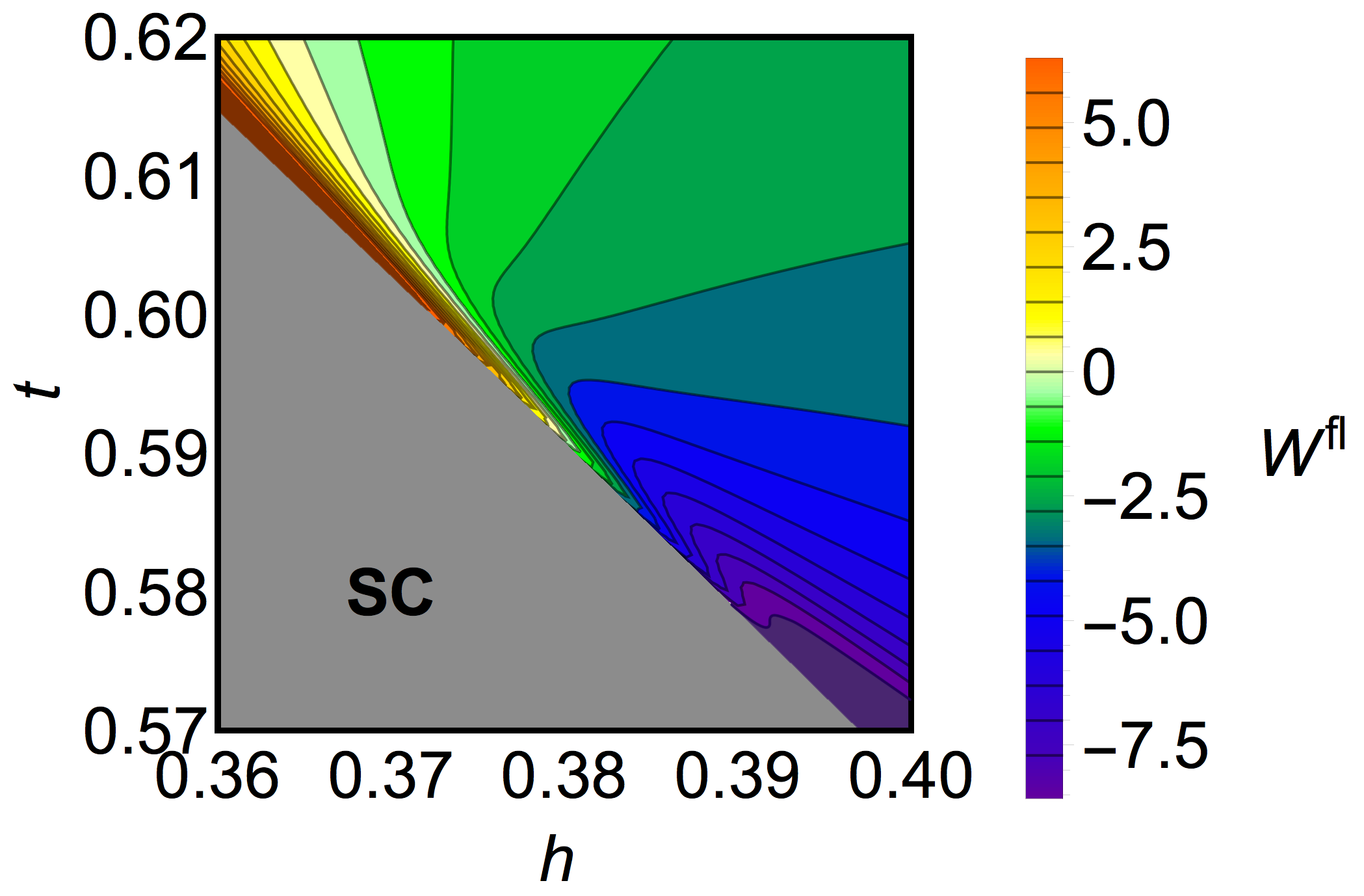}
	\caption{(Color online) \textit{Top:} The temperature and magnetic field dependence of the
		relaxation rate $W^{\rm{fl}}$ in case of a very weak pair-breaking $\protect%
		\gamma _{\protect\phi }=0.003$. The thick isoline (red) represents a zero relaxation rate, while the dashed isolines correspond to relaxation rate values of $-1$ and $-2$.
		The mesh-line $t^*$ (red) marks the critical temperature for $\protect\gamma _{\protect\phi }\to 0$, while the light (cyan)
		contour line indicates the value of $W^{\rm{fl}}$ at $h_{\rm{c2}}(t^\ast)$ ($-3.04$) (see Fig.~%
		\protect\ref{fig.2Dcrit}). \textit{Bottom:} Contour plot of the region close to $h_{\mathrm{c2}}(t^*)$.}
	\label{fig.NMR100}
\end{figure}

\begin{figure}[tbp]
	\includegraphics[width=1.0\columnwidth]{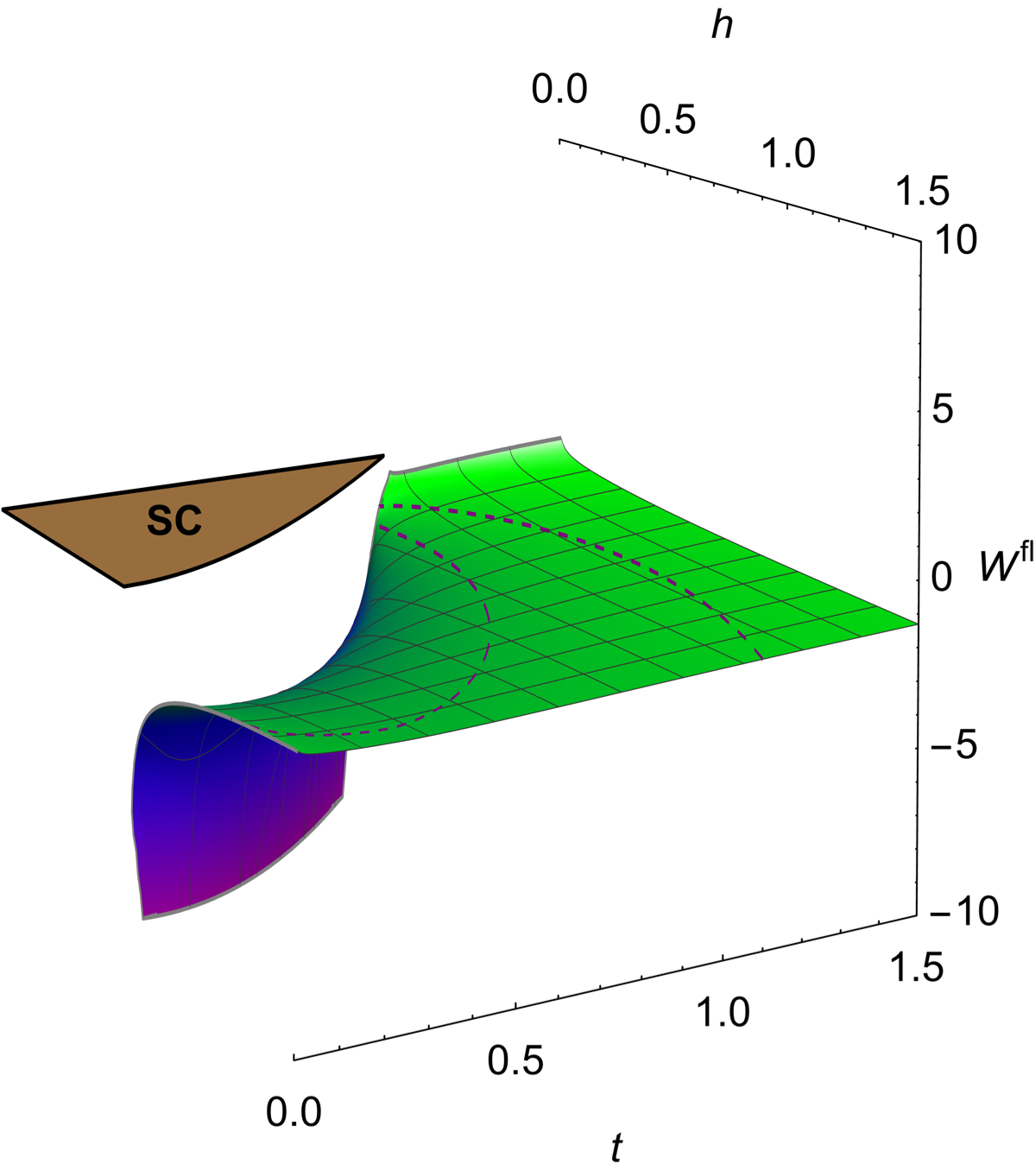}
	\caption{(Color online) The temperature and magnetic field dependence of the relaxation rate $W^{\rm{fl}}$ in case of a strong pair-breaking $\protect\gamma _{\protect\phi }=0.3$. The dashed isolines correspond to relaxation rate values of $-1.5$ and $-2$.}
	\label{fig.NMR1}
\end{figure}

The results of numerical analysis of Eq.~(\ref{TOT}) for different
pair-breaking rates are presented in Figs.~\ref{fig.NMR100}--\ref{fig.NMR1}. In the case of a small enough pair-breaking, there is a large domain of the phase diagram where
superconducting fluctuations result in the increase of the NMR relaxation rate
(see Fig.~\ref{fig.NMR100}). Growth of the pair-breaking suppresses MT contribution and when ${\gamma _{\phi }\sim 1}$ the only effect of quasi-particle DOS suppression
on $W^{\rm{fl}}$ dominates in the whole phase diagram (see Fig.~\ref{fig.NMR1}). It is
interesting that even in the absence of the pair-breaking (${\gamma _{\phi
}\rightarrow 0}$) there exists a crossover temperature $T^{\ast }_0$
below which the MT relaxation process is suppressed by strong magnetic fields and the fluctuation correction $W^{\rm{fl}}$ cannot be positive. In the case of a two-dimensional superconductor $T_{0,\rm{2D}}^{\ast }\approx 0.6T_{\mathrm{c0}}.$ The temperature and field dependence of $W^{\rm{fl}}(T,H)$ near the point $\left\{ T_{\rm{2D}}^{\ast },H_{\rm{c2}}\left( T_{\rm{2D}}^{\ast }\right)\right\} $ is very singular, see the close-up view of its vicinity in the bottom panel of Fig.~\ref{fig.NMR100}.

\begin{figure*}[tbp]
	\includegraphics[width=\textwidth]{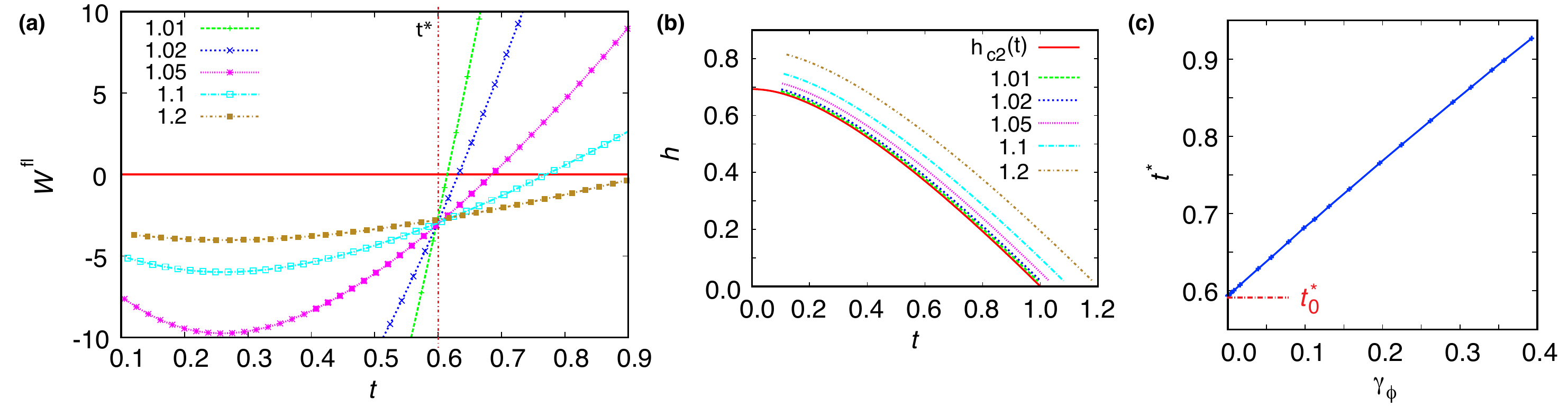}
	\caption{(Color online) Total correction (\textbf{a}) in the 2D case for $\protect\gamma_{\protect\phi }=0.003$ calculated parallel to, but at different distances from the line $h_{\mathrm{c2}}(t)$ in the fluctuation regime. Different colors correspond to different separations defined by the scaling parameter $a$ from $h_{\mathrm{c2}}(t)$ given in the legend, see the (\textit{b}) panel. (\textit{c}) $t^{\ast}(\gamma_{\phi})$ shows a linear dependence for all experimentally relevant values of $\gamma_{\phi}$. For fixed temperatures below $t^{\ast}$ the relaxation rate $W^{\rm fl}$ is monotonically increasing with field, while for larger temperatures the relaxation rate is non-monotonic and grows strongly when approaching $h_{c2}$ from above. The asymptotic value is $t^{\ast}_0=t^{\ast}(\gamma_{\phi}=0)=0.59$.}\label{fig.2Dcrit}
\end{figure*}

Closer analysis of the crossover region, see Fig.~\ref{fig.2Dcrit}, reveals that near $h_{\rm{c2}}(t^{\ast })$ the total correction curves calculated along the line $h_{\rm{c2}}(t)$ cross at a single point at temperature $t^\ast(\gamma_\phi)$, i.e. the total correction to the NMR relaxation rate becomes independent of the field close to the point ${\{t^\ast, h_{\rm{c2}}(t^\ast)\}}$. This can also be seen on Fig.~\ref{fig.NMR100}, where the isoline corresponding to the value ${W^{\rm{fl}}(t^\ast, h_{\rm{c2}}(t^\ast)) \approx -3.04}$ is seen to be parallel to the ${t = t^\ast}$ line in the immediate vicinity of the superconducting region.

Below the crossover temperature $t^\ast$, the total correction exhibits monotonic (increasing) field-dependence for fixed temperature ${t < t^\ast}$. For ${t \ll h_{\rm{c2}}(t)}$, both in the regime of quantum and thermal fluctuations, our numerical analysis is in full agreement with the asymptotic expressions~(\ref{quant}) and (\ref{therm}), confirming the negative sign of the total correction. At the same time, Fig.~\ref{fig.2Dcrit} reveals a non-monotonic behavior at intermediate temperatures ${t \lesssim t^\ast}$ when going along the $h_{\rm{c2}}(t)$ line.

Above the crossover temperature, the field dependence of $W^{\rm{fl}}$ always shows a non-monotonic behavior as a result of the two competing contributions, as can be seen by from Fig.~\ref{fig.NMR100}. The total correction is positive (for not-too-strong pair-breaking $\gamma _{\phi}$) close to the line $h_{\rm{c2}}(t)$; it then decreases rapidly reaching a minimum negative value at some intermediate distance from $h_{\rm{c2}}(t)$ before increasing up to zero when sufficiently far from the superconducting region.

Overall, our result for the total fluctuation correction $W^{\rm{fl}}$ is in qualitative agreement with that of Ref.~\onlinecite{ERS99}, where the temperature range between $0.75T_{\rm{c0}}$ and $1.5T_{\rm{c0}}$ was analyzed to compare with experimental data. The authors correctly point out the strong dependence of $W^{\rm{fl}}$ on the pair-breaking parameter $\gamma_\phi$.  Our analysis based on Eq.~(\ref{TOT}) in the entire temperature range along $H_{\rm{c2}}(T)$ enables us to identify the temperature $T^{\ast}(\gamma_\phi)$ at which the DOS and MT relaxation mechanisms fully compensate each other, such that the fluctuation correction ${W^{\rm{fl}}}$ completely vanishes (in the leading order of perturbation theory). The dependence of this temperature on the pair-breaking parameter $\gamma_\phi$ is presented in Fig.~\ref{fig.2Dcrit}c). The asymptotic crossover temperature $T^\ast_0$ is then defined as $T^{\ast}(0)$, i.e. the temperature below which the negative DOS contribution always dominates, regardless of the values of $\gamma_\phi$ and $h$. The physical picture behind this observation is explained in more detail in Section~VII.

\section{Quasi-two-dimensional superconductor}

\begin{figure}[tbp]
\includegraphics[width=1.0\columnwidth]{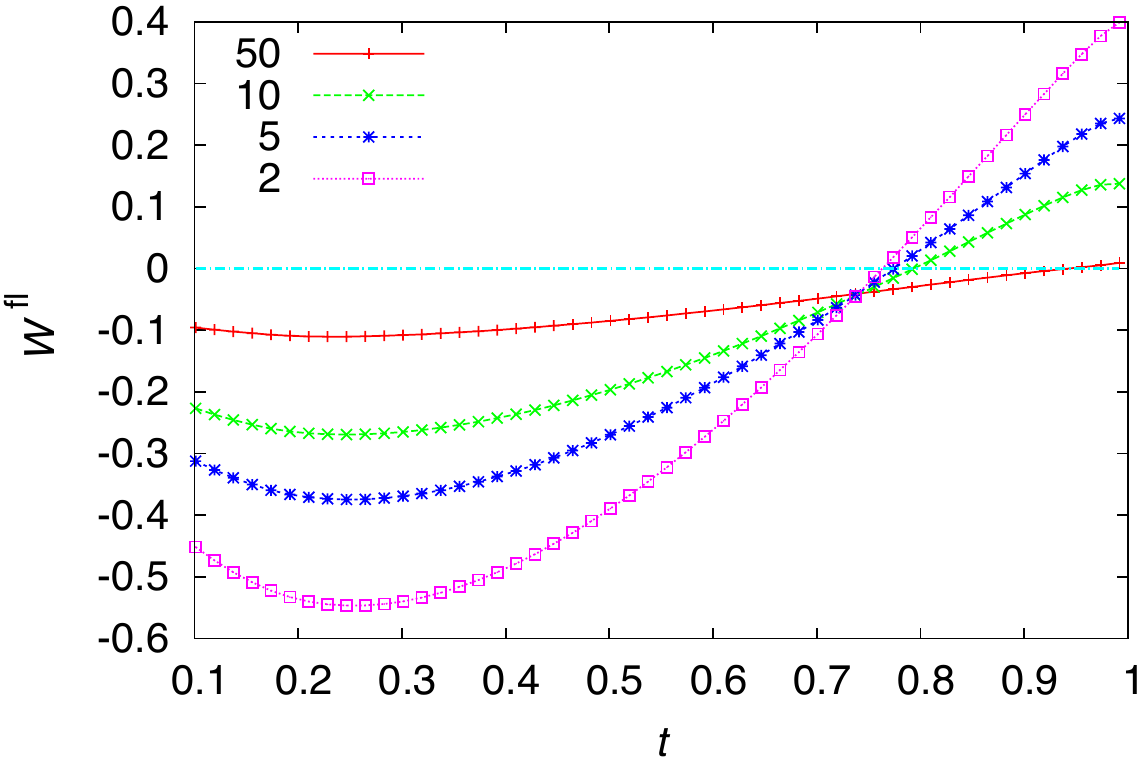}
\caption{(Color online) Quasi-2D result for different $r$ values very close
to $h_{\mathrm{c2}}(t)$ with $a=1.01$ (see Fig.~\ref{fig.2Dcrit}) for $\protect\gamma_{\protect\phi }=0.003$.}
\label{fig.3Dcrit}
\end{figure}

The effect of three-dimensionality of the spectrum can be easily accounted
for by the direct generalization of the Eq.~(\ref{TOT}). Indeed, the
properties of a quasi-two dimensional superconductor can be described well
in the framework of the phenomenological Lawrence-Doniach (LD) model\cite%
{LD70}, which provides a Ginzburg-Landau functional for a layered
superconductor. In the case under consideration, when the magnetic field is
applied perpendicular to the layers, it takes the form: 
\begin{eqnarray*}
\mathcal{F}^{(\rm{LD})}\!\left[ \Psi \right] &=&{\sum_{l}}\!\int\! d^{2}r\!\left[\! \alpha T_{\mathrm{c0}}\epsilon \left\vert \Psi _{l}\right\vert ^{2} \!+\! \frac{b}{2}%
\left\vert \Psi _{l}\right\vert ^{4}\right. \\
&&\left. +\frac{1}{4m}\left\vert \left( \mathbf{\nabla }_{\parallel } \!-\! 2ie \mathbf{A}_{\parallel }\right) |\Psi _{l}\right\vert ^{2} \!+\! \mathcal{J}
\left\vert \Psi _{l+1}-\Psi _{l}\right\vert ^{2}\right] .
\end{eqnarray*}%
Here $\Psi _{l}$\ is the order parameter of the $l$-th superconducting layer
and the phenomenological constant $\mathcal{J}$ is proportional to the
energy of the Josephson coupling between adjacent planes. The gauge with ${A_{z}=0}$ is chosen.

In the immediate vicinity of $T_{\mathrm{c}}$, the LD functional is reduced
to the GL one with the effective mass ${M=\left(4\mathcal{J}s^{2}\right)^{-1}}$ along $z$%
-direction, where $s$ is the inter-layer spacing. One can relate the value
of $\mathcal{J}$ to the coherence length along the $z$-direction, or, what
is more convenient, with the degree of three-dimensionality of the system ${r=4\xi _{z}^{2}/s^{2}}$: ${\mathcal{J}=\alpha T_{\mathrm{c0}}r/2}$ (see Ref.~\ocite{LV09} pp.~26 and 220). Corresponding generalization can be done also
in the microscopic approach: it is enough to enrich the propagator and
Cooperons (or directly the final Eq.~(\ref{TOT})) by the account of the
transversal motion: ${\omega _{\mathrm{c}}\left(m+\frac{1}{2}\right)\rightarrow \omega _{%
\mathrm{c}}\left(m+\frac{1}{2}\right)+\frac{\mathcal{J}}{2}(1-\cos q_{z}s)}$ and perform
the additional integration over the transversal momentum. In order to avoid
the cumbersome expressions, let us show explicitly how it works only for the
regular DOS contribution~(\ref{T_1reg}): 
\begin{widetext}
\begin{equation}
W^{\rm{DOS}(\rm{LD},\mathrm{reg})}(t,h)=\frac{AN(0)}{8\pi
^{2}\xi ^{2}}h\int_{-\pi /s}^{\pi /s}\frac{%
dq_{z}}{2\pi }\sum_{m=0}^{M}\sum_{k=0}^{\infty }\frac{\psi ^{\prime \prime }%
\left[ \frac{1+|k|}{2}+\frac{2h}{t}\frac{\left( 2m+1\right) }{\pi ^{2}}+%
\frac{4r}{\pi ^{2}t}\sin ^{2}\frac{q_{z}s}{2}\right] }{\ln t+\psi \left[ 
\frac{1+|k|}{2}+\frac{2h}{t}\frac{\left( 2m+1\right) }{\pi ^{2}}+\frac{4r}{%
\pi ^{2}t}\sin ^{2}\frac{q_{z}s}{2}\right] -\psi \left( \frac{1}{2}\right) }.
\label{3D}
\end{equation}%
\end{widetext}
One can see that the averaging in Eq.~(\ref{3D}) over the transversal modes
effectively reduces it to the same Eq.~(\ref{T_1reg}) with addition of some
positive constant $\lambda r/t$ in the arguments of all polygamma functions.
Hence, in order to satisfy the same conditions for ${T^{\ast }\left(
r=0\right)}$ but at some temperature ${T^{\ast }\left( r>0\right)}$, one
should have%
\begin{equation*}
\frac{h_{\mathrm{c2}}\left( t_{\rm{2D}}^{\ast }\right) }{t_{\rm{2D}}^{\ast }}=\frac{%
h_{\mathrm{c2}}\left( t_{r}^{\ast }\right) }{t_{r}^{\ast }}+\pi ^{2}\lambda \frac{r}{%
t_{r}^{\ast }}.
\end{equation*}%
This means that ${h_{\mathrm{c2}}\left( t_{\rm{2D}}^{\ast }\right)/t_{\rm{2D}}^{\ast
}>h_{\mathrm{c2}}\left( t_{r}^{\ast }\right) /t_{r}^{\ast }}$, or, taking into account
the monotonous increase of the second critical field with the decrease of
temperature, one makes sure that ${t_{r}^{\ast }>t_{\rm{2D}}^{\ast }}$, i.e. the
temperature $T^{\ast }$ should grow with the increase of $r.$ This
qualitative speculation is confirmed by the numerical study of
correspondingly generalized Eq.~(\ref{TOT}) (see Fig.~\ref{fig.3Dcrit}), where the asymptotic crossover temperature is increased to ${t^{\ast}_{0,\rm{3D}}\approx 0.75}$.

\section{Discussion}

Here we discuss the physical aspect of the results obtained and consequences
for the general understanding of the fluctuation picture. As already
explained above, it is the MT process of the self-electron pairing at the
self-intersecting trajectories that is responsible for the growth of $W(T,H).
$ Its contribution to the NMR relaxation rate is proportional to the
superconducting interaction strength $g\left( T,H\right) $ and to the total
probability for the formation of such trajectories (see Ref.~[\onlinecite{LV09}]):%
\begin{equation*}
W_{\rm{2D} }^{\rm{MT}}(T,H)\sim g( T,H) \int_{\xi _{\rm{FCP}}\left(
T,H\right) /v_{F}}^{\min \{\ell _{\phi },L_{H}\}/v_{F}}\frac{\lambda
_{F}v_{F}dt}{Dt}.
\end{equation*}%
Here $L_{H}=\sqrt{c/2eH}$ is the FCP magnetic length, while ${\xi_{\rm{FCP}}( T,H)}$ is its effective size. Close to $T_{\mathrm{c0}}$
and in weak fields, where the long wave-length Ginzburg-Landau fluctuation
picture takes place ${\xi _{\rm{FCP}}( T,H) =\xi _{\rm{GL}}=\xi _{\rm{xy}}/\sqrt{\epsilon }}$. In the opposite case of quantum fluctuations at zero
temperature and in the vicinity of $H_{\mathrm{c2}}(0)$, the size of FCP
clusters, including many pairs, is of the order of ${\xi _{\rm{xy}}/\sqrt{%
\widetilde{h}}}$, but the length corresponding to one of them is ${\xi_{\rm{FCP}}(0,H) \sim \xi _{\rm{xy}}}$ (see Refs.~[\onlinecite{GVV11a}] and [\onlinecite{GVV11b}]). At some intermediate region of temperatures along the
line $H_{\mathrm{c2}}(T)$ to low temperatures the increasing magnetic field
\textquotedblleft breaks\textquotedblright\ GL waves and the vortex
description becomes more adequate. The obtained crossover temperature $%
T^{\ast }$ allows us to define where it happens: namely for ${\xi _{\rm{FCP}}\left[
T^{\ast },H( T^{\ast }) \right] \sim L_{H}=\sqrt{c/2eH(T^{\ast }) }}$, where the MT mechanism of NMR relaxation becomes irrelevant when going to lower temperatures (see Figs.~\ref{fig.NMR100} and \ref{fig.2Dcrit}). The crossover temperature depends on the pair-breaking parameter $\gamma _{\phi }$ and becomes minimal in the limit ${\gamma _{\phi
}\rightarrow 0}$ with a value ${t^{\ast }_0\approx 0.6}$ which can be clearly
seen in Figs.~\ref{fig.NMR100} and \ref{fig.NMR1}. 

Now let us return to discussion of the experiments of Ref.~[\onlinecite{LR,LRZ05}], which partially motivated this work. The authors observed a well
pronounced peak of $W(T,H)$ versus magnetic field in the low temperature
part ($t=0.12;0.25)$ when crossing the line $H_{\mathrm{c2}}(T)$ and
attributed it to the possible manifestation of the quantum fluctuations.
Unfortunately, our analysis of all fluctuation contributions definitely
excludes this hypothesis: below ${t^{\ast }_0\approx 0.6}$ fluctuations can only open
the spin gap in the NMR relaxation rate, but they cannot lead to its growth.
The observed decay above $H_{\mathrm{c2}}(T)$ is therefore not related to
quantum fluctuations.

\section{Acknowledgements}

We express our deep gratitude to A. Rigamonti and A. Lasciafari for
attracting our attention to their experiments and numerous elucidating
discussions. A.V. was partially supported by the U.S. Department of Energy,
Office of Science, Office of Advanced Scientific Computing Research and
Materials Sciences and Engineering Division, Scientific Discovery through
Advanced Computing (SciDAC) program.

\appendix

\section{Numerical evaluation of the NMR relaxation rate}

In order to utilize the complete expression for the NMR relaxation rate $%
W^{\rm{fl}}$ to analyze experimental data we need an efficient and accurate
method to evaluate Eq.~(\ref{TOT}) numerically. Here we describe the method
used throughout this work. The integral contributions ($z$-integrations) can
be straight-forwardly evaluated using a suitable quadrature scheme. Here we
use the Gauss-Legendre 5-point method, which also allows integration of
integrable poles or principle values. Due to the presence of the $%
\sinh^{-2}(\pi z)$ term in the integrand we can restrict the support to $z\in%
[-5,5]$. Outside this interval the integrand is smaller than the numerical
accuracy (of double precision floating point numbers). This sum over
Landau-levels is calculated up to $M=(tT_{\mathrm{c0}}\tau)^{-1}$ explicitly.

In contrast, the summation over $k$ in the MT contribution to Eq.~(\ref{TOT}%
) is more involved and only slowly converging. For the numerical summation
of the $k$-sum we separate the $k=0$ term and sum from $k=1$ to $k_{\max }$
(twice, due to symmetry) which is determined by the arguments of the $%
\mathcal{\psi }^{(n)}$ functions being equal to $\Omega = 1000$. For $k\geq
k_{\max }$ we transform the sum into an integral and use only the asymptotic
expressions for the polygamma functions as the difference to the exact
expression is again below the floating point accuracy. Then the integration
variable is inverted and we have a finite integral for the remaining part of
the sum.

Therefore we concentrate on 
\begin{equation*}
S^{\rm{MT}}_m\equiv \sum_{k=-\infty }^{\infty }\frac{\mathcal{E}_{m}^{\prime
\prime }\left( t,h,\left\vert k\right\vert \right) }{\mathcal{E}_{m}\left(
t,h,\left\vert k\right\vert \right) }
\end{equation*}
and write

\begin{eqnarray*}
S_{m}^{\rm{MT}} &=&\left[ \sum_{k=0}^{k_{\max }-1}\left( 2-\delta _{0,k}\right)
+2\int_{k_{\max }}^{\infty }dk\right] \frac{\mathcal{E}_{m}^{\prime \prime
}\left( t,h,\left\vert k\right\vert \right) }{\mathcal{E}_{m}\left(
t,h,\left\vert k\right\vert \right) } \\
&\equiv &S_{m}^{\rm{MT}(s)}+S_{m}^{\rm{MT}(i)}
\end{eqnarray*}

with
\begin{equation*}
k_{\max }=\max \left\{ 2\Omega -\left\lfloor \frac{4h}{\pi ^{2}t}\left(
2m+1\right) \right\rfloor ,1\right\}\,.
\end{equation*}

The sum part $S^{\rm{MT}(s)}_m$ is calculated straightforwardly, which leaves
the calculation of the ``rest-integral'' $S^{\rm{MT}(i)}_m$:

\begin{eqnarray*}
S_{m}^{\rm{MT}(i)} &=&\frac{1}{2}\int\limits_{k_{\max }}^{\infty }\!\!dk\,\frac{\psi ^{\prime\prime }\left( \frac{1 + k}2+x_{m}\right) }{\ln t-\psi \left( \frac{1}{2}\right)
+\psi \left( \frac{1 + k}2+x_{m}\right) } \\
&\circeq&\! -\frac{1}{2}\!\int\limits_{k_{\max }}^{\infty}\!\!dk\,\frac{\left( \frac{1 + k}2+x_{m}\right)^{-2}}{\ln t-\psi \left( \frac{1}{2}\right) -\ln
(2)+\ln \left( 1+k+2x_{m}\right) }
\end{eqnarray*}

with $x_{m}\equiv \frac{2h}{t}\frac{\left( 2m+1\right) }{\pi ^{2}}$.

A convenient substitution is

\begin{align*}
&\frac1z \!=\!\frac{8}{\pi ^{2}}+\frac{(1+k)t}{h\left( m+1/2\right) }=\frac{8}{\pi
^{2}x_{m}}\left[ x_{m}+\frac{1+k}2\right]\,,  \\
&\frac{dz}{z^{2}} =-\frac{t}{h\left( m+1/2\right) }dk=-\frac{4}{\pi ^{2}}%
\frac{dk}{x_{m}}\,, \\
&z_{\max } =\frac{\pi ^{2}}{4}\left( 2+\frac{1+k_{\max}}{x_{m}}\right) ^{-1}\,.
\end{align*}

Therefore, 
\begin{eqnarray*}
S_{m}^{\rm{MT}(i)} \!&=&\! \frac{\pi ^{2}}{8}\int\limits_{z_{\max }}^{0}\frac{dz}{z^{2}}\!%
\frac{x_{m}\left( \frac{8z}{\pi ^{2}x_{m}}\right) ^{2}}{\ln \left( t\pi
^{2}x_{m}/4\right) -\psi \left( \frac{1}{2}\right) -\ln (2)-\ln (z)} \\
&=&-\frac{8}{\pi ^{2}x_{m}}\int\limits_{0}^{z_{\max }}dz\frac{1}{A_{m}-\ln z} \\
&=&-\frac{2t}{h\left( m+1/2\right) }\int\limits_{0}^{z_{\max }}dz\frac{1}{A_{m}-\ln
z}
\end{eqnarray*}

with $A_{m}\equiv \ln \left[ h\left( m+1/2\right) \right] -\psi \left( \frac{%
1}{2}\right) -\ln (2)$.

This integral is integrable and calculated by the Gauss-Legendre 5-point
method (with avoids the singular point at $z=0$) with only a few support
points in the small interval $0$ to $z_{max}$ using $125$ support points.

Overall this yields a highly accurate numerical value of the $k$-sums.

In the quasi-two-dimensional case the additional finite $q$-integral is
calculated by the Gauss-Legendre 5-point method using $25$ support points,
which is sufficient to obtain high accuracy.


\bibliographystyle{apsrev4-1}
\bibliography{NMR.bib}

\end{document}